\documentclass[aps,prl,showpacs,twocolumn,floats,epsfig]{revtex4}
\usepackage{amssymb}
\usepackage{amsbsy}
\usepackage{amsmath}
\usepackage{epsfig}

\begin{document}

\title {Confinement-deconfinement transition in a generalized Kitaev model.}

\author{S. Mandal$^{(1)}$, S. Bhattacharjee$^{(2)}$,
K. Sengupta$^{(3,4)}$ and R. Shankar$^{(1)}$}

\affiliation{$^{(1)}$ The Institute of Mathematical Sciences, C.I.T
Campus, Chennai-600113, India.\\
$^{(2)}$ CCMT, Department of Physics, Indian Institute of Science, Bangalore-560012, India.\\
$^{(3)}$ Theoretical Physics Division, Indian Association for the
Cultivation of Sciences, Jadavpur, Kolkata-700032, India.\\
$^{(4)}$ TCMP division, Saha Institute of Nuclear Physics, 1/AF
Bidhannagar, Kolkata-700064, India.}

\date{\today}

\begin{abstract}

We present a spin model, namely, the Kitaev model augmented by a
loop term and perturbed by an Ising Hamiltonian and show that it
exhibits both confinement-deconfinement transitions from spin liquid
to antiferromagnetic/spin-chain/ferromagnetic phases and topological
quantum phase transitions between gapped and gapless spin liquid
phases. We develop a Fermionic mean-field theory to chart out the
phase diagram of the model and estimate the stability of its spin
liquid phases which might be relevant for attempts to realize the
model in optical lattices. We also conjecture that some of the
confinement-deconfinement transitions in the model, predicted to be
first order within the mean-field theory, may become second order
via a defect condensation mechanism.

\end{abstract}


\maketitle

Quantum phase transition from ordered to paramagnetic phases in
two-dimensional (2D) spin models has been a subject of recent
interest\cite{subir1}. Such paramagnetic phases and associated
quantum phase transitions have been conjectured to be of relevance
to the properties of several strongly correlated systems including
cuprates and quasi-2D organic materials \cite{lee1,
senthil1,subir2}. A class of these paramagnetic phases which do not
break any constituent symmetries of their underlying lattice are
called spin-liquids and are generally believed to be natural
candidates for paramagnetic phases obtained by disordering
non-collinear spin-ordered magnets \cite{subir3}. However, the
precise criteria for realization of spin liquids and the nature of
quantum phase transitions to them from spin-ordered phases are far
from being settled issues \cite{senthil2}. The physics of these
spin-liquid phases can be described by representing the spins in
terms of spinons, which are Fermionic $CP(N)$ fields, coupled to
bosonic gauge fields \cite{lee1,subir3}. In most commonly studied
examples, the symmetry group associated with these gauge fields are
either $U(1)$ or $Z_2$; the corresponding spin-liquids being dubbed
as $U(1)$ or $Z_2$ spin liquids. In the ordered phase of spins, the
spinons are confined while the spin-liquid paramagnets constitute
phases with gapped or gapless deconfined spinon excitations. Thus
the transition between these phases serve as examples of
confinement-deconfinement transitions for spinons. Examples of such
transitions has been studied in several high energy and condensed
matter models \cite{fradkin1,fradkin2,nagaosa1}.

\begin{figure}
 \rotatebox{0}{\includegraphics*[width=.8\linewidth]{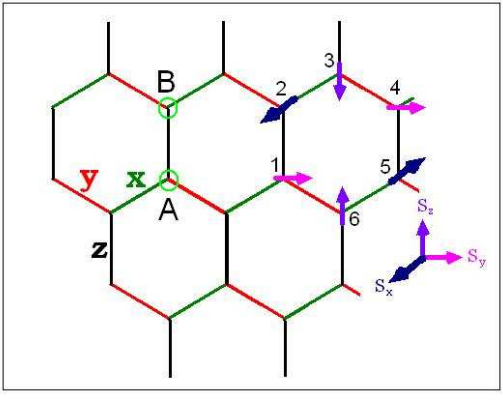}}
\caption{(Color online) Schematic representation of the Kitaev model
on a honeycomb lattice showings the different links $x$, $y$ and $z$
and the two sublattices A and B. The sites labeled $1..6$ and their
spin configuration is a schematic representation for a classical
configuration with $\langle W_p \rangle=1$ } \label{fig1}
\end{figure}

In this letter, we present a spin model, namely the Kitaev model
augmented by a loop term and perturbed by an Ising model, on a 2D
hexagonal lattice with the Hamiltonian
\begin{eqnarray}
H &=& H_K + H_L + H_I, \quad H_K = -\sum_{j \in A} \sum_{\alpha\,
link} J_{\alpha}
\sigma_{j}^{\alpha} \sigma_{j_{\alpha}^{'}}^{\alpha} \nonumber\\
H_I &=& \lambda J \sum_j \sum_{all \, links} \sigma_j^z
\sigma_{j'}^z, \quad H_L = -\kappa J \sum_p W_p
\end{eqnarray}
where  $j_{\alpha}^{'}$ is the nearest neighbor of $j$ connected by
the $\alpha=x,y,z$ link of the hexagonal lattice as shown in Fig.\
\ref{fig1}, $\sigma^{\alpha}$ are the usual Pauli matrices, $W_p=
\sigma_1^x \sigma_2^y \sigma_3^z \sigma_4^x \sigma_5^y \sigma_6^z$
is the loop operator, where, as shown in Fig.\ \ref{fig1}, $1..6$
denotes sites of a hexagonal plaquette, $A$ and $B$ denotes two
sublattices of the honeycomb lattice, and $j'$ denotes all nearest
neighbors of $j$. We demonstrate that this model exhibits
confinement-deconfinement transitions from deconfined spin liquid to
confined Ising ordered antiferromagnetic (AFM), ferromagnetic (FM)
or spin-chain (SC) phases and can therefore serve as a test bed to
study such transitions. In addition, we show that the model also
supports two distinct spin liquid phases with gapped and gapless
deconfined spinon excitations and exhibits a topological quantum
phase transition between them. To the best of our knowledge, the
model presented here is the only spin model which supports spin
liquid phases with both gapped and gapless deconfined spinon
excitations and displays both confinement-deconfinement and
topological quantum phase transitions. We chart out the phase
diagram of this model using a Fermionic mean-field theory and
estimate the stability of the deconfined phases as a function of the
strengths of the loop and Ising terms. There have been proposals for
experimentally realizing the Kitaev model in systems of ultracold
atoms and molecules trapped in optical lattices \cite{expt1}. Any
such physical realization of this model will always have
contaminating interactions; our work thus provides an estimate of
the stability of its gapless spin liquid phase against such
interactions. Finally, we conjecture that the
confinement-deconfinement transition between the gapped deconfined
Kitaev phase and the confined Ising AFM or FM phases, predicted to
be first order within the mean-field theory, may turn out to be
second order via a defect condensation mechanism at large $\kappa$.

The Kitaev model with Hamiltonian $H_K$ is a rare example of a 2D
spin model which can be exactly solved
\cite{kit1,feng1,chen1,baskaran1,ks1}. The ground state of the
model, for the parameter regime $|J_1-J_2|\le J_3 \le J_1+J_2$,
supports a gapless phase \cite{kit1} which can be described as a
$Z_2$ spin liquid constituting a Fermi sea of deconfined gapless
spinons and static $Z_2$ gauge fields. These spinons become gapped
beyond this regime. Both the gapped and the gapless spinon phases
have unit expectation value for a non-local loop operator $W_p$:
$\langle W_p \rangle = 1$. They are distinguished by the value of the
Chern number which is $\pm 1$ for the gapless phase and $0$ for the
gapped phases.

In the present work, we study the effect of augmenting the Kitaev
model with a loop term and perturbing it with an Ising Hamiltonian.
In what follows, we shall study the Kitaev model in the isotropic
limit and scale all energies by $J$: $J_{\alpha}=J=1$ and obtain the
phase diagram of $H$ as functions of dimensionless couplings
$\kappa$ and $\lambda$. We note at the outset, that the ground state
of the system at $\lambda=0$ is the Kitaev ground state with gapless
deconfined Fermionic spinons coupled to a frozen gauge field
configuration. This ground state is exactly captured by the
Fermionic theory which makes it an ideal starting point for our
analysis. The ground state of the system at $\lambda \to \pm \infty$
is the Ising AFM/FM which are magnetically ordered phases with
confined spinons. Thus the model, by construction, clearly supports
confinement-deconfinement transitions.

We begin by mapping the spin model $H$ to a Fermionic model $H_F$,
by using the Jordan-Wigner transformations \cite{chen1}. We choose a
path $\{i(n)\}$ which runs along the $x$ and $y$ bonds and define
$\sigma_{j(n)}^{z}\equiv ic_{j(n)} b^z_{j(n)}$, $\sigma_{j(2n)}^x
\equiv c_{j(2n)}\prod_{m<2n}\sigma_{k(m)}^z$, $\sigma_{j(2n+1)}^x
\equiv b^z_{j(2n+1)}\prod_{m<2n+1}\sigma_{k(m)}^z$, and $\sigma^y_j=
i\sigma^x_j\sigma^z_j$, where $c_j,~b^z_j$ are Majorana fermions
operators at site $j$ [$(j(2n),j(2n-1))$ are the $x$ bonds and
$(j(2n+1),j(2n))$ are the $y$ bonds]. The resultant Hamiltonian
becomes
\begin{eqnarray}
H_F &=& - \sum_{j\in A} \Big[ \sum_{\alpha=x,y\, links } i c_{j}
c_{j_{\alpha}^{'}} + \sum_{z \,link}i b^z_{j} c_j i b^z_{j_z^{'}}
c_{j_z^{'}} \Big]
\nonumber \\
&& -\kappa \sum_{j,k \in plaquette} \sum_{z link} i b_j^z
b_{j_z^{'}}^z i b_k^z b_{k_z^{'}}^z \nonumber\\
&& + \lambda \sum_j \sum_{\alpha = all \,links} i b^z_{j} c_j i
b^z_{j_{\alpha}^{'}} c_{j_{\alpha}^{'}} \label{maj1}
\end{eqnarray}
where the subscript $j,k \in plaquette$ indicates that the sum is
over sites which belong to the $A$ sublattice of a given plaquette
as schematically shown in Fig.\ \ref{fig1}. Note that for
$\lambda=0$, the operators $ib^z_jb^z_{j^\prime_z}$, commute with
the Hamiltonian and are therefore a constants of motion. In this
limit, $H$ is exactly solvable. When $\lambda$ is turned on, these
operators acquire dynamics and their fluctuations are ultimately
expected to confine the spinons through a confinement-deconfinement
transition.

To make further progress, we treat $H_F$ within an RVB type
mean-field theory \cite{lee1} and introduce the mean-fields on the
sites (corresponding to spin ordering) and on links (corresponding
to the emergent gauge fields) of the hexagonal lattice: $\langle i
b_j^z c_j\rangle = \langle \sigma^z_j \rangle = \Delta_{1(2)}$,
$\langle i b^z_j c_{j_{\alpha}^{'}}\rangle = \beta_{\alpha}$,
$\langle i b^z_j b^z_{j_{\alpha}^{'}}\rangle = \gamma_{\alpha}$, and
$\langle i c_j c_{j_{\alpha}^{'}}\rangle = \gamma_{0\alpha}$. Note
that keeping in mind the bipartite nature of the hexagonal lattice
and to allow for possible AFM phases, we have introduced two
mean-fields $\Delta_1$ and $\Delta_2$ corresponding to the two
sublattices shown in Fig.\ \ref{fig1}. In terms of these
mean-fields, one can decompose the quartic terms in $H_F$ as
\begin{eqnarray}
i b_j^z b_{j_z^{'}}^z i b_k^z b_{k_z^{'}}^z &=& i \gamma_z (b_j^z
b_{j_z^{'}}^z + b_k^z b_{k_z^{'}}^z) - \gamma_z^2 \nonumber\\
i b^z_{j} c_j i b^z_{j_{\alpha}^{'}} c_{j_{\alpha}^{'}} &=&
\gamma_{\alpha} i c_j c_{j_{\alpha}^{'}} + \gamma_{0\alpha}i b^z_{j}
b^z_{j_{\alpha}^{'}} - \gamma_{\alpha} \gamma_{0\alpha} \nonumber\\
&& - \Delta_1 i b^z_{j_{\alpha}^{'}} c_{j_{\alpha}^{'}} - \Delta_2 i
b^z_{j} c_j + \Delta_1 \Delta_2 \nonumber\\
&& - i \beta_{\alpha} ( b^z_{j} c_{j_{\alpha}^{'}} +
b^z_{j_{\alpha}^{'}} c_j ) - \beta_{\alpha}^2,
\end{eqnarray}
where we have only considered mean-fields which are either on-site
or reside on links between two neighboring sites. The resultant
quadratic mean-field Hamiltonian, in momentum space, can be written
as
\begin{eqnarray}
H_{\rm mf} &=& \frac{1}{N} \sum_{\vec k} \Big[ J_0 \left( \alpha +
e^{i k_1} +
e^{i(k_1+k_2)} \right) c_{\vec k}^{A \dagger} c_{\vec k}^{B} \nonumber\\
&& +J_0^{'} \left( \beta -2 \kappa \gamma_z /J_0^{'} + e^{i k_1} +
e^{i
(k_1+k_2)} \right) b_{\vec k}^{A \dagger} b_{\vec k}^{B} \nonumber\\
&& + (i c_{\vec k}^{A \dagger} b_{\vec k}^{B}- i b_{\vec k}^{A
\dagger} c_{\vec k}^{B} ) \left( \beta_z (1+\lambda) + \beta_x
e^{ik_1} \right.
\nonumber\\
&& \left. + \beta_y e^{i k_2} \right)- c_1 b_{\vec k}^{A \dagger}
c_{\vec k}^{A} - c_2 b_{\vec k}^{B
\dagger} c_{\vec k}^{B} + \rm {h.c} \Big]  \nonumber\\
&& + \kappa \gamma_z^2 - (1+\lambda) \gamma_z \gamma_{0z} + (1+3
\lambda) \Delta_1 \Delta_2 \label{mf2}
\end{eqnarray}
where $J_0= (1+\lambda \gamma_{x})$, $\alpha J_0 =
\left(1+\lambda\right) \gamma_z$, $J_0^{'}= \lambda \gamma_{0x})$,
$\beta J^{'}_0 = \left(1+\lambda\right) \gamma_{0z}$, $c_{1(2)}=
(1+3\lambda) \Delta_{1(2)}$, and the momentum $\vec k = k_1 \hat e_1
+ k_2 \hat e_2$ with the unit vectors $\hat e_1 = \hat x + \hat
y/\sqrt{3}$ and $\hat e_2 = 2\hat y/\sqrt{3}$.
\begin{figure}
 \rotatebox{0}{\includegraphics*[width=.8\linewidth]{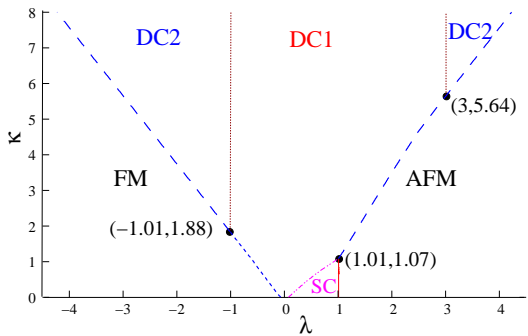}}
\caption{(Color online) The mean-field phase diagram for the model.
The blue dashed lines represent confinement-deconfinement
transitions which may become second order via defect condensation
mechanism. The triple points occur at
$(\lambda^{\ast}_1,\kappa^{\ast}_1)=(1.01,1.07)$ and
$(\lambda^{\ast}_2,\kappa^{\ast}_2)=(3,5.64)$ for $\lambda>0$ and
$(\lambda^{\ast}_3,\kappa^{\ast}_3)=(-1.01,1.88)$ for $\lambda <0$.
} \label{fig2}
\end{figure}

We now minimize $H_{\rm mf}$ numerically and obtain the mean-field
phase diagram of the model as a function of $\lambda$ and $\kappa$.
Note that the mean field solution is exact at $\lambda=0$. This
phase diagram is shown in Fig.\ \ref{fig2}. In accordance with our
earlier expectation, we find that at large positive (negative)
$\lambda$, the ground state of the model is an Ising AFM (FM) which
corresponds to confined phase of spinons while at small $\lambda$,
the model exhibits a deconfined gapless phase DC1. The transition
between these two phases at low $\kappa$ and for positive $\lambda$
occurs via a SC phase, which corresponds to AFM alignment of spins
along chains in $x$ direction of the hexagonal lattice with
ferromagnetic arrangement of such chains in the $y$ direction. For
negative $\lambda$, there is a direct transition to the FM phase. At
high enough values of $\kappa$, we find another gapped deconfined
phase DC2 and second order topological quantum phase transitions
between DC1 and DC2 phases. The confinement-deconfinement
transitions at high $\kappa$ always occur from DC2 to AFM/FM phases.
These transitions are predicted to be first order within mean-field
theory. The phase diagram exhibits two triple points at
$(\lambda^{\ast}_1, \kappa^{\ast}_1)=(1.01,1.07)$ and
$(\lambda^{\ast}_2, \kappa^{\ast}_2)=(3,5.64)$ for $\lambda > 0$.
These represent meeting points of AFM, SC and DC1 and AFM, DC2 and
DC1 phases respectively. For $\lambda <0$, there is one triple point
$(\lambda^{\ast}_3, \kappa^{\ast}_3)=(-1.01,1.88)$ where the FM,
DC1, and DC2 phases meet. We also note that our mean-field analysis
also gives an estimate for the stability of the deconfined phase of
the Kitaev model ($-0.07 \le \lambda_c \le 0.08$ for $\kappa=0$)
under external perturbing Ising term which may be important for
physical realization of the Kitaev model and for quantum computing
proposals based on it \cite{expt1,topcomp1}.

\begin{figure}
 \rotatebox{0}{\includegraphics*[width=.8\linewidth]{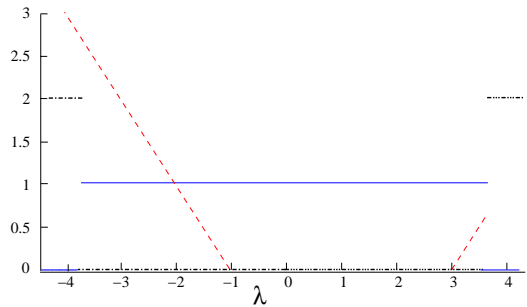}}
\caption{(Color online) Plot of the loop order parameter (solid blue
line), the spinon gap (red dashed line) and the FM and the AFM order
parameters (black dash-dotted lines) as a function of $\lambda$ for
$\kappa=7$. } \label{fig3}
\end{figure}

The plot of the loop order parameter $\langle W_p \rangle$, the
spinon gap, and the AFM and the FM order parameters as obtained from
the mean-field theory, is shown, for $\kappa=7$, as a function of
$\lambda$ in Fig.\ \ref{fig3}. We note that all the order parameters
show discontinuous changes at the transition points indicating first
order transitions. The spinon gap, in contrast, increases linearly
and continually with $\lambda$ indicating a second order quantum
phase transition between DC1 and DC2 phases. The presence of this
topological quantum phase transition and the linear variation of the
spinon gap with $\lambda$ can be understood qualitatively from
$H_{\rm mf}$. For large $\kappa$, it requires a large $\lambda$ to
destabilize the Kitaev ground state in favor of Ising AFM/FM. In
addition, numerically we find that in the Kitaev phase $\gamma_z
(\gamma_x) \sim 1(0)$. As a result, beyond a critical value of
$\lambda=\lambda_c$, the effective couplings along the links,
$J_{1,2} \sim (1+ \lambda \gamma_x)$, $J_3 \sim
\gamma_z(1+\lambda)$, fail to satisfy $|J_1-J_2|\le J_3 \le J_1+J_2$
thus leading to a gapped phase via a topological quantum phase
transition \cite{kit1,feng1}. The spinon gap in this gapped phase
varies linearly with $J_3$ \cite{feng1,ks1} and hence shows a linear
variation on $\lambda$. At small $\kappa$, the
confinement-deconfinement transitions to the SC/FM phases occur
before $\lambda_c$ is reached and hence the topological phase
transitions do not occur.

\begin{figure}
 \rotatebox{0}{\includegraphics*[width=0.65\linewidth]{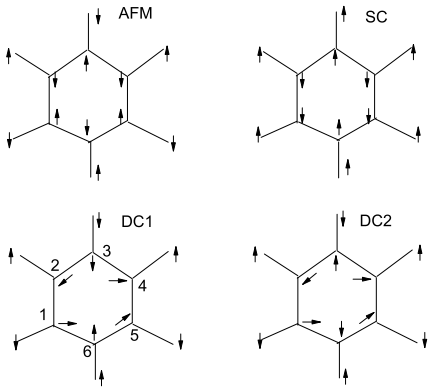}}
\caption{Schematic representation of classical spin configuration of
the AFM phase and possible defects that can destabilize this phase.
See text for details.} \label{fig4}
\end{figure}

Next, we argue that it is possible that some of the
confinement-deconfinement transitions, predicted to be first order
by mean-field theory, can become second order in the large $\kappa$
limit via a defect condensation mechanism \cite{senthil3}. We note
that similar second-order transitions have been seen in numerical
studies of related models \cite{vidal1, nishimori1}. We begin with
the case $\lambda > 0$ and start from the Ising AFM ground state.
The possible classical defect spin configurations over this ground
states whose condensation can lead to either the Kitaev or the SC
phases are shown in Fig.\ \ref{fig4}. The energy of these defects
are given by $E_{sc} = 8(\lambda -1) $, $E_{DC1} = 14 \lambda - 6 -
\kappa$, and $E_{DC2} = 10 \lambda - 2 - \kappa$. Note that
$E_{AFM}=E_{DC1}=E_{sc}=E_{DC2}=0$ at $\lambda=\lambda^{\ast}=1$ and
$\kappa=\kappa^{\ast} = 8$ which represents a triple point in the
$\kappa-\lambda$ plane. For the SC phase to occur, we need
$E_{sc}=0$ and $ E_{sc} \le E_{DC1}, E_{DC2}$ which yields the
conditions
\begin{eqnarray}
\lambda_c^{sc} &=& 1 \quad \kappa_c^{sc} \le 6\lambda +2, \quad
\kappa_c^{sc} \le 2 \lambda +6 \label{condsc}
\end{eqnarray}
Similar analysis lead to conditions for instability due to
condensation of the defects DC1 and DC2:
\begin{eqnarray}
\lambda_c^{DC1} &=& \frac{\kappa_c^{DC1}}{14} + \frac{3}{7} \quad
\lambda_c^{DC1} \le 1, \quad \kappa^{DC1} \ge  6 \lambda +2,
\label{conddc1} \\
\lambda_c^{DC2} &=& \frac{\kappa_c^{DC2}}{10} + \frac{1}{5} \quad
\lambda_c^{DC2} \ge 1, \quad \kappa^{DC2} \ge  2 \lambda +6.
\label{conddc2}
\end{eqnarray}

From Eqs.\ \ref{condsc}, \ref{conddc1}, and \ref{conddc2}, we find
that the SC phase destabilize the AFM phase at $\lambda=1$ for all
$\kappa \le \kappa^{\ast}$. This represents an order-order
transition which is expected to be first order. As we decrease
$\lambda$ further below $1$, the deconfined gapless Kitaev phase
takes over at some point. The nature of this transition is not
predictable from this condensation mechanism, since the transition
do not take place from the AFM phase. For $\kappa \ge
\kappa^{\ast}$, the AFM phase is destabilized in favor of the
deconfined phase DC2 when $\lambda^c = \kappa^c/10 +1/5$. The DC2
phase represents a disordered phase in terms of the spin variables
and thus this transition can be second order. For $\kappa
> \kappa^{\ast}$, the energy of DC1 becomes lower than that of DC2
for $\lambda \le 1$. This indicates a second order transition
between the two deconfined phases with no broken symmetries and is
therefore a the topological quantum phase transition. Note that this
transition line is independent of $\kappa$ and therefore is expected
to appear as a vertical line in the $\kappa-\lambda$ plane. A
similar discussion for $\lambda <0$ shows that the defect
condensation energies of the phases DC1 and DC2 (the SC phase is not
energetically favorable for $\lambda < 0$) over the FM state are
$E'_{DC1} = 14 |\lambda| + 6 - \kappa$ and $E'_{DC2} = 10 |\lambda|
+ 2 - \kappa$. We find that such a mechanism works only for large
$\kappa$ where $E'_{DC1}$ and $E'_{Dc2}$ can be negative and in this
regime the FM phase is always destabilized by the DC2 phase leading
to a confinement-deconfinement at $|\lambda| = \kappa/10 - 1/5$. For
lower $\kappa$, we expect to find a transition between the DC1 and
the FM phase which is not mediated by such defect condensation. The
phase diagram obtained by this simple qualitative discussion of the
defect condensation mechanism described above has most of the major
features shown in Fig.\ \ref{fig2} and a direct AFM to DC1
transition line at intermediate $\kappa$ for $\lambda>0$) and this
gives us some confidence about it's qualitative correctness
\cite{comment1}. A quantitatively accurate study of the obtained
phase diagram requires a more sophisticated treatment of quantum
fluctuations and is left for future study.

To conclude, we have presented a spin model, namely the Kitaev
model, augmented by a loop term and perturbed by an Ising
Hamiltonian, and have shown that the model exhibits a rich phase
diagram with both confinement-deconfinement and topological quantum
phase transitions. We have estimated that the topological phase of the Kitaev
model is unstable to about 10\% contamination by Ising interactions.


\begin{thebibliography}{99}
\bibitem{comment1} The two triple points predicted by mean-field
theory for $\lambda>0$ merge in this description. We expect this
artifact to be cured by inclusion of perturbative quantum
corrections to defect condesation energies.
\end{thebibliography}
\end{document}